\documentstyle[12pt]{article}
\def\hybrid{\topmargin -20pt	\oddsidemargin 0pt
	\headheight 0pt	\headsep 0pt
	\textwidth 6.25in	% A4 paper
	\textheight 9.5in	% A4 paper
	\marginparwidth .875in
	\parskip 5pt plus 1pt	\jot = 1.5ex}

\hybrid

\catcode`\@=11

\def\marginnote#1{}
\newcount\hour
\newcount\minute
\newtoks\amorpm
\hour=\time\divide\hour by60
\minute=\time{\multiply\hour by60 \global\advance\minute by-\hour}
\edef\standardtime{{\ifnum\hour<12 \global\amorpm={am}%
	\else\global\amorpm={pm}\advance\hour by-12 \fi
	\ifnum\hour=0 \hour=12 \fi
	\number\hour:\ifnum\minute<10 0\fi\number\minute\the\amorpm}}
\edef\militarytime{\number\hour:\ifnum\minute<10 0\fi\number\minute}
\def\draftlabel#1{{\@bsphack\if@filesw {\let\thepage\relax
   \xdef\@gtempa{\write\@auxout{\string
      \newlabel{#1}{{\@currentlabel}{\thepage}}}}}\@gtempa
   \if@nobreak \ifvmode\nobreak\fi\fi\fi\@esphack}
	\gdef\@eqnlabel{#1}}
\def\@eqnlabel{}
\def\@vacuum{}
\def\draftmarginnote#1{\marginpar{\raggedright\scriptsize\tt#1}}

\def\draft{\oddsidemargin -.5truein
	\def\@oddfoot{\sl preliminary draft \hfil
	\rm\thepage\hfil\sl\today\quad\militarytime}
	\let\@evenfoot\@oddfoot	\overfullrule 3pt
	\let\label=\draftlabel
	\let\marginnote=\draftmarginnote
   \def\@eqnnum{(\theequation)\rlap{\kern\marginparsep\tt\@eqnlabel}%
\global\let\@eqnlabel\@vacuum}  }

\def\preprint{\twocolumn\sloppy\flushbottom\parindent 2em
	\leftmargini 2em\leftmarginv .5em\leftmarginvi .5em
	\oddsidemargin -.5in	\evensidemargin -.5in
	\columnsep .4in	\footheight 0pt
	\textwidth 10.in	\topmargin  -.4in
	\headheight 12pt \topskip .4in
	\textheight 6.9in \footskip 0pt
	\def\@oddhead{\thepage\hfil\addtocounter{page}{1}\thepage}
	\let\@evenhead\@oddhead	\def\@oddfoot{}	\def\@evenfoot{} }

\def\numberbysection{\@addtoreset{equation}{section}
	\def\theequation{\thesection.\arabic{equation}}}

\def\underline#1{\relax\ifmmode\@@underline#1\else
	$\@@underline{\hbox{#1}}$\relax\fi}

\def\titlepage{\@restonecolfalse\if@twocolumn\@restonecoltrue\onecolumn
     \else \newpage \fi \thispagestyle{empty}\c@page\z@
	\def\thefootnote{\fnsymbol{footnote}} }

\def\endtitlepage{\if@restonecol\twocolumn \else \newpage \fi
	\def\thefootnote{\arabic{footnote}}
	\setcounter{footnote}{0}}  %\c@footnote\z@ }

\catcode`@=12
\relax

\def\figcap{\section*{Figure Captions\markboth
	{FIGURECAPTIONS}{FIGURECAPTIONS}}\list
	{Figure \arabic{enumi}:\hfill}{\settowidth\labelwidth{Figure 999:}
	\leftmargin\labelwidth
	\advance\leftmargin\labelsep\usecounter{enumi}}}
 \relax
\def\tablecap{\section*{Table Captions\markboth
	{TABLECAPTIONS}{TABLECAPTIONS}}\list
	{Table \arabic{enumi}:\hfill}{\settowidth\labelwidth{Table 999:}
	\leftmargin\labelwidth
	\advance\leftmargin\labelsep\usecounter{enumi}}}
 \relax
\def\reflist{\section*{References\markboth
	{REFLIST}{REFLIST}}\list
	{[\arabic{enumi}]\hfill}{\settowidth\labelwidth{[999]}
	\leftmargin\labelwidth
	\advance\leftmargin\labelsep\usecounter{enumi}}}
 \relax
\makeatletter
\newcounter{pubctr}
\def\publist{\@ifnextchar[{\@publist}{\@@publist}}
\def\@publist[#1]{\list
	{[\arabic{pubctr}]\hfill}{\settowidth\labelwidth{[999]}
	\leftmargin\labelwidth
	\advance\leftmargin\labelsep
	\@nmbrlisttrue\def\@listctr{pubctr}
	\setcounter{pubctr}{#1}\addtocounter{pubctr}{-1}}}
\def\@@publist{\list
	{[\arabic{pubctr}]\hfill}{\settowidth\labelwidth{[999]}
	\leftmargin\labelwidth
	\advance\leftmargin\labelsep
	\@nmbrlisttrue\def\@listctr{pubctr}}}
 \relax
\makeatother
\newskip\humongous \humongous=0pt plus 1000pt minus 1000pt

\newif\ifdtup

\relax

\def\beq{\begin{equation}}
\def\eeq{\end{equation}}
\def\bea{\begin{eqnarray}}
\def\eea{\end{eqnarray}}
\def\bq{\begin{quote}}
\def\eq{\end{quote}}

\def\PL{{\it Phys.Lett.~}}
\def\NP{{\it Nucl.Phys.~}}
\def\PR{{\it Phys.Rev.~}}

\def\MPL{{\it Mod.Phys.Lett.~}}
\def\SPJ{{\it Sov.Phys.-JETP~}}

\def\gappeq{\mathrel{\rlap {\raise.5ex\hbox{$>$}}
{\lower.5ex\hbox{$\sim$}}}}

\def\lappeq{\mathrel{\rlap{\raise.5ex\hbox{$<$}}
{\lower.5ex\hbox{$\sim$}}}}

\def\Toprel#1\over#2{\mathrel{\mathop{#2}\limits^{#1}}}
\def\Pbar{\Toprel{\lower 2pt
\hbox{$\scriptscriptstyle(-)$}}\over{\partial}}

\def\Toprel#1\over#2{\mathrel{\mathop{#2}\limits^{#1}}}
\def\Gbar{\Toprel{\lower 2pt
\hbox{$\scriptscriptstyle(-)$}}\over{G}}

\def\Toprel#1\over#2{\mathrel{\mathop{#2}\limits^{#1}}}
\def\pbar{\Toprel{\lower 2pt
\hbox{$\scriptscriptstyle(-)$}}\over{P}}

\def\Toprel#1\over#2{\mathrel{\mathop{#2}\limits^{#1}}}
\def\Pibar{\Toprel{\lower 2pt
\hbox{$\scriptscriptstyle(-)$}}\over{\Pi}}

\def\thefootnote{\fnsymbol{footnote}}

\begin{document}
\pagestyle{empty}
\begin{flushright}
{CERN-TH.6790/93}
\end{flushright}
\vspace*{5mm}
\begin{center}
{\bf CONSTRUCTION OF STRING SOLUTIONS} \\
{\bf AROUND NON-TRIVIAL BACKGROUNDS} \\
\vspace*{.7cm} {\bf C. Kounnas} \footnote{On leave from Lab. de Physique
Th\'eorique
de l' ENS, F-75231, Paris, FRANCE.}\\
\vspace*{0.3cm}
Theoretical Physics Division, CERN \\
CH - 1211 Geneva 23 \\
\vspace*{1.5cm}
{\bf ABSTRACT} \\ \end{center}
\vspace*{5mm}
\noindent
We present a way of constructing string solutions around non-trivial
gravitational backgrounds. The proposed solutions are constructed using $N
= 4$ superconformal building blocks with $\hat c = 4$. We give two
different and inequivalent realizations of non-trivial four-dimensional
subspaces, and we show the emergence of the $N = 4$ globally defined
superconformal symmetry. The existence of $N = 4$ world-sheet
symmetry stabilizes our solutions and implies in target space a number of
covariantized supersymmetries around  space-time dependent gravitational
and dilaton backgrounds.

\vspace*{1cm}
\begin{center}
{\it Talk given at the }\\
{\it International Workshop on} \\
{\it ``String Theory, Quantum Gravity and Unification of Fundamental
Interactions"} \\
{\it Rome, 21-26 September 1992} \\
{\it and at the} \\
{\it 26th Workshop: ``From Superstrings to Supergravity"}\\
{\it Erice - Sicily, 5-12 December 1992}
\end{center}
\vspace*{1cm}
\begin{flushleft} CERN-TH.6790/93 \\
January 1993
\end{flushleft}
\vfill\eject
%\pagestyle{empty}
%\clearpage\mbox{}\clearpage

\setcounter{page}{1}
\pagestyle{plain}
\begin{center}
{\bf CONSTRUCTION OF STRING SOLUTIONS} \\
{\bf AROUND NON-TRIVIAL BACKGROUNDS}\\
\vspace*{0.5cm}
Costas KOUNNAS\footnote{On leave from Lab. de Physique Th\'eorique
de l' ENS, F-75231, Paris, FRANCE.} \\
Theoretical Physics Division, CERN \\
CH - 1211 Geneva 23 \\
\vspace*{1cm}
ABSTRACT
\end{center}

\parbox[c]{12.7cm}{\small\noindent
We present a way of constructing string solutions around non-trivial
gravitational backgrounds. The proposed solutions are constructed using $N
= 4$ superconformal building blocks with $\hat c = 4$. We give two
different and inequivalent realizations of non-trivial four-dimensional
subspaces, and we show the emergence of the $N = 4$ globally defined
superconformal symmetry. The existence of $N = 4$ world-sheet
symmetry stabilizes our solutions and implies in target space a number of
covariantized supersymmetries around  space-time dependent gravitational
and dilaton backgrounds.
\small}

\vspace*{1cm}
A way to better understand string theories and their induced low-energy
field theories coupled to Einstein gravity is to study classical string
solutions in the presence of non-trivial space-time backgrounds.

There are two ways to proceed in this direction:
\begin{description}
\item[~(i)] The first  is to use a two-dimensional $\sigma$-model where
the non-trivial backgrounds correspond to some two-dimensional
field-dependent coupling constants. The vanishing of the corresponding
$\beta$-functions is then identified with the background field equation of
motion in the target space-time \cite{aaa}.
\item[(ii)] The second approach consists of constructing directly some
underlying non-trivial conformal field theory, and then trying to interpret
the obtained string vacuum in terms of the target space-time backgrounds.
\end{description}

The two methods are useful and complementary. The $\sigma$-mode approach
provides a clear geometric interpretation, but it has the disadvantage of
the $\alpha^{\prime}$-expansion which is valid only when all curvatures and
derivatives on space-time background fields are small. Via the
$\sigma$-model approach one can easily obtain approximate solutions which
are in fact identical to the classical equation of gravity in the presence
of a dilaton, antisymmetric field and some gauged minimally coupled matter.
However, the possible extension of these approximate solutions to exact
string solutions is in general difficult and an unsolved problem at the
present time.

	The two-dimensional conformal field theory approach takes into account
all orders in $\alpha^{\prime}$ automatically and has the main advantage of
providing exact string vacua. The background interpretation of a given
exact string solution is a notion which is ill-defined in general. Indeed,
the notion of  space-time dimensionality and topology breaks down for a
solution which involves highly curved backgrounds, namely when the metric
and/or gauge field curvatures are of the order of the string scale. A
typical example of the space-time dimensional and topological confusion is
that of the $SU(2)$ level $k$ group manifold compactification. For large $k$
(small curvature) the target space is a three-dimensional sphere $S^3$,
$[\frac{SO(4)}{SO(3)} \simeq SO(3) \simeq SU(2)]$. For small $k$ (high
curvature) this background interpretation fails. It is in fact well known
that the $SU(2)_{k=1}$ WZW  model is equivalent to a $c = 1$ conformal system
defined by one free bosonic co-ordinate compactified on a cycle with a
radius $R = \frac{2}{R} = \sqrt{2}$ (self-dual point).

	Naively one may interpret $T_{(R=\sqrt{2})}$ compactification as
one-dimensional space with $S^1$ topology, which is in contradiction to
the three-dimensional interpretation with $S^3$ topology of the
$SU(2)_{k=1}$. This shows that both the dimensionality as well as the
topology of the target space are not well-defined concepts in string
theory. In general, a background interpretation of a given string solution
exists only when the lower Kaluza-Klein excitations have masses much
smaller than the typical string scale $(M_{st} = \alpha '^{-1/2})$.

In this talk, I will present a class of string solutions which is
constructed   by using some $N = 4$ superconformal systems with $\hat c =
4$ as building blocks. In the limit of small curvatures, these solutions
have a non-trivial ten-dimensional background interpretation. Furthermore,
the globally defined underlying $N = 4$ worldsheet symmetry stabilizes these
solutions under string-loop perturbation, and implies some covariantized
space-time sypersymmetry  around a non-trivial gravitational
and dilaton background on the target space.

	More explicitly, we arrange the degrees of freedom of the ten
supercoordinates in three superconformal systems \cite{bb}-\cite{ee}:
\beq
\hat c = 10 = \{ \hat c = 2 \} + \{ \hat c = 4 \}_1 + \{ \hat c = 4\}_2
\label{1}
\eeq
The $\hat c = 2$ subsystem is saturated by two free superfields. In one
variation of our solutions, one of the two free superfields is chosen to be
the time-like supercoordinate and the other to be  one of the nine
space-like supercoordinates. In other variations, both supercoordinates are
Euclidean or even compactified on a one- or two-dimensional torus.

The remaining eight supercoordinates appear in a group of four in $\{\hat c
= 4\}_1$ and $\{\hat c = 4\}_2$. Both $\{\hat c = 4\}_A$ subsystems show an
$N = 4$ superconformal symmetry of the Ademollo et al. type \cite{ff}. The
non-triviality of our solutions follows from the fact that some
realizations of the $\hat c = 4, N = 4$ superconformal systems exist which
are based on geometrical and topological non-trivial spaces other than the
$T^4/Z_2$ orbifold and the $K_3$ compact Calabi-Yau space. I will
now present  two different realizations.

\vspace*{0.5cm}\noindent
{\bf (A) - $W_k^{(4)}$, semi-wormhole realization} \cite{ggg},\cite{hh}
$\hat c [W_k^{(4)}] = 4$

It is based on a supersymmetric version of the $\{U(1)\times SU(2)_k\}$ WZW
model. The three bosonic coordinates parametrize the $SU(2)_k$ group
manifold while the fourth one is a free field with background charge $Q$.
The four fermionic coordinates are free. In order to obtain a $\hat
c[U(1)\times SU(2)_k] = 4$ for any value of $k$, it is necessary to
balance the central charge deficit of the $SU(2)_k$ by a central charge
benefit of the $U(1)$ background charge $Q$
\bea
\hat c[SU(2)_k] = \frac{2}{3} [3-\frac{6}{k+2} + \frac{3}{2}] = 3 -
\frac{4}{k+2} \nonumber \\
\hat c[U(1)_Q] = \frac{2}{3} [1 + 3Q^2 + \frac{1}{2}] = 1 +
2Q^2
\label{2}
\eea
(The contributions $\frac{3}{2}$ and $\frac{1}{2}$ inside $[\ldots ]$ in the
first and second line are those of the 3+1 free fermions).

{}From Eq. (\ref{2}) one has
$$
\hat c[SU(2)_k] + \hat c[U(1)_Q] = 4 + 2(Q^2 - \frac{2}{k+2})~,
$$
and so $\hat c [ W_k^{(4)}] = 4$ only if
\beq
Q = \sqrt{\frac{2}{k+2}}
\label{3}
\eeq

The existence of $N = 4, \hat c = 4$ superconformal symmetry with this
value of $Q$ is found in Ref. \cite{ggg}. What is extremely interesting is
the background interpretation of the $W_k^{(4)}$ space in terms of a
four-dimensional (semi)-wormhole space given by Callan, Harvey and
Strominger in Ref. \cite{hh}. Indeed, for large $k$, the three $SU(2)_k$
coordinates define a three-dimensional subspace with a non-trivial topology
$S^3$, while the fourth coordinate with a background charge corresponds to
the scale factor of the $S^3$ sphere.

Another interesting interpretation of the $W_k^{(4)}$ space is its
connection with four-dimensional dilaton-axion instantons
\cite{hh}-\cite{kk}. Here, one assumes a string solution based on a
six-dimensional compact manifold $K^6$ with $\hat c = 6$; our
four-dimensional space-time (after Euclidean rotation) is saturated by
$W_k^{(4)}$ with $\hat c = 4$.

The explicit realization of the $N = 4$ symmetry of $W_k^{(4)}$, as well as
some other properties of string solutions based on it, will be discussed
later after I present a new and inequivalent realization of a $\hat c = 4, N
= 4$ superconformal system.

\vspace*{0.5cm}\noindent
{\bf {B} - $\Delta_k^{(4)}$ Cigar-Bell and Trumpet-Bell realizations}
\cite{ee} $\hat c [\Delta_k^{(4)}] = 4$

For large values of $k$ (small curvatures), the $\Delta_k^{(4)}$ has a
four-dimensional interpretation which is different from $W_k^{(4)}$. The
underlying two-dimensional action associated to $\Delta_k^{(4)}$ is an exact
superconformal theory based on a supersymmetric version of a gauged WZW
model, namely,
\beq
\Delta_k^{(4)} \equiv \bigg\{\bigg [\frac{SU(2)}{U(1)}\bigg]_k \times \bigg[
\frac{SL(2,R)}{U(1)}\bigg]_{k+4}\bigg\}_{\rm SUSY}
\label{4}
\eeq
The relation among the levels $k$ and $k' = k+4$ ensures
$\hat c[\Delta^{(4)}_k] = 4$ for any $k$
\bea
\hat c[\frac{SU(2)}{U(1)}]_{ss} = \frac{2}{3} [\frac{3k}{k+2} -1+1]
= 2 - \frac{4}{k+2} \nonumber \\
\hat c[\frac{SL(2,R)}{U(1)}]_{ss} = \frac{2}{3} [\frac{3k'}{k'-2} -1+1]
= 2 + \frac{4}{k'-2}
\label{5}
\eea
and so $\hat c[\Delta^{(4)}_k] = 4$ when $k' = k+4$. As we will show
later, the above relation among the levels is necessary for the existence
of an $N = 4$ superconformal symmetry of Ademollo type \cite{ff}.

For large $k$, the $\Delta_k^{(4)}$ space has a dimensional target space
interpretation based on a non-trivial background metric $G_{ij}$ and
dilaton field $\Phi$ given by \cite{lll}-\cite{rr},\cite{bb}-\cite{ee}
$$\phantom{a non-trivial}
ds^2 = k\bigg\{(d\alpha )^2 + \tan^2\alpha d\theta^2\bigg\} + k'
\bigg\{(d\beta )^2  + \tanh^2\beta d\varphi^2\bigg\}\phantom{interpretation}
(6a) $$
$$\phantom{a non-trivial}
2\Phi = \log \cos^2\alpha + \log \cosh^2\beta + {\rm const.}
\phantom{dimensional target  interpretation}
(6b)
$$
with
$$
\alpha \in [0,\frac{\pi}{2}] \bigcup [\pi ,\frac{3}{2}\pi ]~,~~\beta\in
[0,\infty ]~,~~\theta , \varphi \in [0,2\pi]
$$
The term proportional to $k$ in Eq. (6a) parametrizes the two-dimensional
subspace defined by the $\bigg [\frac{SU(2)}{U(1)}\bigg]_k$ parafermionic
theory \cite{ss} and the one proportional to $k'$ is the two-dimensional
subspace which is defined by the non-compact parafermionic theory
\cite{tt},\cite{uu},\cite{mm} based on the
$\bigg[\frac{SL(2,R)}{U(1)}\bigg]_{k'}$ axial gauged WZW model. It is well
known that a different metric $\tilde G_{ij}$ and dilaton function
$\tilde\Phi$ are obtained if one chooses a vector gauging instead of the
axial one \cite{pp},\cite{nn}:
$$
\matrix{\phantom{interpretation  on }(U(1)^V\times U(1)^V) \rightarrow
d\tilde s^2 = &~~k\bigg\{  (d\alpha )^2 + \frac{1}{\tan^2\alpha}
d\theta^2\bigg\}  \phantom{ based a non-trivial}\cr
\phantom{interpretation  on } &+ k'\bigg\{ (d\beta
)^2 + \frac{1}{\tanh^2\beta} d\varphi^2\bigg\}
\phantom{ a non-trivial}(7a)\cr}
$$
$$
\phantom{(U(1)^V\times U(1)^V) \rightarrow} 2\tilde\Phi = ~~\log\sin^2\alpha
+ \log\sinh^2\beta + {\rm const.}\phantom{interpretation based on a the} (7b)
$$
\addtocounter{equation}{2}
In both versions of gaugings, $\Delta_k^{(4)}$ has always one non-compact
coordinate $(\beta )$ and three compact ones $(\alpha ,\theta ,\varphi )$.
$(G_{ij},\Phi )$ and $(\tilde G_{ij},\tilde\Phi )$ are related by a
generalized duality transformation \cite{oo}:
\beq
R(t)\rightarrow \frac{1}{R(t)}~,~~\Phi\rightarrow\Phi + \log R(t)
\label{8}
\eeq
For later purposes, it is more convenient to use the complex notation:
$$
z = (\sin\alpha )e^{i\theta}~, ~~\omega = (\sinh\beta )e^{i\varphi}~,~~
({\rm axial~case})
$$
\beq
{\rm  and}\phantom{For later purpose, it is more convenient to use
the complex notation:} \eeq
$$
\tilde z = (\cos\alpha )e^{i\theta}~,~~ y = (\cosh\beta )e^{i\varphi}~,
{}~~({\rm vector~case})
\label{9}
$$
In terms of $z$ and $w$, $G_{ij}$ and $\Phi$ are given by
\beq
(ds)^2 = k~\frac{dzd\bar z}{1-z\bar z} + k'~\frac{dwd\bar w}{w\bar w + 1}~,
\label{10}
\eeq
and

$$2\Phi = \log (1-z\bar z ) + \log (w\bar w + 1) + {\rm const.}$$.

The dual metrics $\tilde G_{ij}$ and $\tilde\Phi$ are given in terms of
$\tilde z$ and y as:
$$
(d\tilde s)^2 = k~\frac{d\tilde zd\tilde z*}{1-\tilde z\tilde z*} +
k'~\frac{dyd\bar y}{y\bar y - 1}~,
$$
\beq
{\rm and} \phantom{For later purpose, it is more convenient to use
the complex notation:}
\eeq
$$
2\tilde\Phi  = \log (1-\tilde z\tilde z* ) + \log (y\bar y - 1)~.
\label{11}
$$

{}From Eqs. (\ref{10}) and (11), one observes that the
$\bigg(\frac{SU(2)}{U(1)}\bigg)_k$ metric and dilaton part are self-dual,
 $(z$ and $\tilde z$), while the
$\bigg(\frac{SU(2,R)}{U(1)}\bigg)_{k'}$ is not. The $w$ subspace is regular
while the $y$ subspace is singular).

	The $z$ subspace (or $\tilde z$) defines a two-dimensional Bell. Its
metric $G_{z\bar z}$, the Ricci tensor $R_{z\bar z}$ and its scalar
curvature $R^{(z)}$ are singular at the boundary of the Bell $(z = 1)$:
\beq
G_{z\bar z} = \frac{k}{1-z\bar z}~,~~R_{z\bar z} = \frac{-1}{(1-z\bar
z)^2~}, ~~R^{(z)} =  \frac{-1}{k(1-z\bar z)}~,
\label{12}
\eeq
The $z$-Bell is negatively curved for any value of $\vert z\vert < 1$.

	The $w$ subspace is regular everywhere with finite positive curvature and
is asymptotically flat for $\vert w\vert \rightarrow\infty$
\beq
G_{w\bar w} = \frac{k'}{w\bar w+1}~,~~R_{w\bar w} = \frac{+1}{(w\bar
w+1)^2}~, ~~R^{(w)} =  \frac{+1}{k'(w\bar w+1)}~,
\label{13}
\eeq
It has a cigar shape with maximal curvature at $w = 0$.

The $y$ subspace has a   different shape from its $w$-dual. It looks like a
two-dimensional trumpet with infinite curvature at the boundary $(y = 1)$.
It is  positively curved everywhere $(\vert y\vert >1)$ and is
asymptotically flat for $\vert y\vert\rightarrow\infty$:
\beq
G_{y\bar y} = \frac{k'}{y\bar y-1}~,~~R_{y\bar y} = \frac{+1}{(y\bar
y-1)^2}~, ~~R^{(y)} =  \frac{+1}{k'(y\bar y-1)}~,
\label{14}
\eeq

	We have therefore two different versions of the $\Delta_k^{(4)}$ space.
The first one is that of the $(z,w)$-BELL-CIGAR four-dimensional space and
the second one is that of the $(\tilde z,y)$ BELL-TRUMPET four-dimensional
space.

Up to now we have discussed the geometrical structure of the bosonic
coordinates of the $\Delta_k^{(4)}$ space. In order to complete the
description of our solution, we must include the fermionic superpartners of
$(z,w)$ coordinates, e.g., four Weyl Majorana left-handed two-dimensional
fermions $(\psi^a_+, a = 1,2,3,4)$ as well as four Weyl-Majorana
right-handed ones $(\psi^a_-, a = 1,2,3,4)$.

	Because of the $N = 1$ local supersymmetry, the interactions among
fermions are fixed in terms of the two-dimensional $\sigma$-model
backgrounds $G_{ij}, B_{ij}$ and $\Phi$). In the superconformal gauge, one
has the following generic form for the $N = 1$ $\sigma$-model action (in
the absence of $B_{ij} = 0$):
\beq
S = -\frac{1}{2\pi} \int d\xi d\bar\xi \bigg\{ V_+^aV_-^a - \frac{1}{2}
(\psi^a_+\nabla_-\psi^a_+ - \psi^a_-\nabla_+\psi^a_-)
-\frac{1}{2} R_{ab,cd} \psi^a_+\psi^b_+\psi^c_-\psi^d_- +\Phi R^{(2)}\bigg\}
\label{15}
\eeq
where $a = 1,2,3,4$ are local flat indices and
\beq
V_+^a = E^a_i\partial X^i~,~~V^a_- = E_i^a \bar\partial X^i~,~~{\rm
with}~~ G_{ij} = E_i^aE^a_j
\label{16}
\eeq
The $\nabla_-$ and $\nabla_+$  denote the left- and right-handed covariant
derivatives acting on left- and right-handed fermions $\psi^a_+ ,
\psi^a_-$:
\bea
\nabla_-\psi^a_+ = \bar\partial\psi_+^a +
\Gamma^a_{\phantom{a}bc}\psi^c_+V^b_- \nonumber \\
\nabla_+\psi^a_- = \partial\psi_-^a +
\Gamma^a_{\phantom{a}bc}\psi^c_-V^b_+
\label{17}
\eea
Since $\Delta_K^{(4)}$ is defined as a direct product of two
two-dimensional subspaces, the $\sigma$-model action can be written as:
\beq
S(\Delta_K^{(4)}) = S\bigg[\frac{SU(2)}{U(1)}\bigg]_k +
S\bigg[\frac{SL(2,R)}{U(1)}\bigg]_{k'}
\label{18}
\eeq
Because of the identity
\beq
R_{ab,cd}\psi^a_+\psi^b_+\psi^c_-\psi^d_- = 2R
\psi^1_+\psi^2_+\psi^1_-\psi^2_-~,
\label{19}
\eeq
valid in both subspaces and thanks to the relations \cite{ee}
$$
\matrix{
\phantom{  thanks to relations}
-(R^{(z)} + \Gamma_zG^{z\bar z}\Gamma_{\bar z}) &=& \frac{1}{k}~;~~
\bigg(\frac{SU(2)}{U(1)}\bigg)_k
&\phantom{valid in both subspaces}(20a)\cr
\phantom{  thanks to relations}
-(R^{(w)} + \Gamma_wG^{z\bar z}\Gamma_{\bar w}) &=& \frac{-1}{k'}~;~~
\bigg(\frac{SL(2,R)}{U(1)}\bigg)_{k'}
&\phantom{valid in both subspaces}(20b)}
$$
\addtocounter{equation}{1}
\noindent
it is possible to rewrite the $\sigma$-model action in a more convenient
form  which shows in particular (at least at the classical level) that
the fermions can be described in terms of two free bosonic fields
compactified in a special radius. Indeed, using Eqs. (19) and (20), one
finds \cite{ee}:
\bea
S\bigg[\frac{SU(2)}{U(1)}\bigg]_k =
\frac{-1}{4\pi} \int d\xi d\bar\xi
&&\bigg\{ \partial A\bar\partial A + t(A)^2
(\partial\theta_A - \sqrt{\frac{2}{k}}~\psi^1_+\psi_+^2)~
(\bar\partial\theta_A - \sqrt{\frac{2}{k}}~\psi^1_-\psi^2_-)\nonumber \\
&&-(\psi^1_+\bar\partial\psi^2_+ + \psi^2_+\bar\partial\psi^2_+ +
\psi^1_-\partial\psi^1_- + \psi^2_-\partial\psi^2_-)\nonumber \\
&& +\frac{2}{k} (\psi^1_+\psi^2_+)~(\psi^1_-\psi^2_-) + \log
C^2(A)R^{(2)}\bigg\}
\label{21}
\eea
where $A, \theta_A$ are rescaled fields so that, in the large $k$ limit,
real bosonic fields are conventionally normalized [see Eq. (9)]:
\bea
&&\sin^2~\frac{A}{\sqrt{2k}} = \sin^2\alpha = z\bar z \nonumber \\
&&-i~\frac{1}{2}~\log~\frac{z}{\bar z} = \theta = \frac{\theta_A}{\sqrt{2k}}
\nonumber \\
&&t(A)^2 = \frac{z\bar z}{1-z\bar z} = \tan^2~\frac{A}{\sqrt{2k}}\nonumber \\
&&C(A)^2 = 1-z\bar z = \cos^2~\frac{A}{\sqrt{2k}}
\label{22}
\eea
In a similar way, the
\bea
S\bigg[\frac{SL(2,R)}{U(1)}\bigg]_{k'} =
\frac{-1}{4\pi} \int d\xi d\bar\xi
&&\bigg\{ \partial  B\bar\partial B + T(B)^2
(\partial\theta_B - \sqrt{\frac{2}{k'}}~\psi^3_+\psi_+^4)~
(\bar\partial\theta_B - \sqrt{\frac{2}{k'}}~\psi^3_-\psi^4_-)\nonumber \\
&&-(\psi^3_+\bar\partial\psi^4_+ + \psi^4_+\bar\partial\psi^4_+ +
\psi^3_-\partial\psi^3_- + \psi^4_-\partial\psi^4_-)\nonumber \\
&& -\frac{2}{k'} (\psi^3_+\psi^4_+)~(\psi^3_-\psi^4_-) + \log
C(B)^2R^{(2)}\bigg\}
\label{23}
\eea
where $B$ and $\theta_B$ are now defined in terms of $w$ fields:
\bea
&&\sinh^2~\frac{B}{\sqrt{2k'}} = \sin^2\beta = w\bar w \nonumber \\
&&-i~\frac{1}{2}~\log~\frac{w}{\bar w} = \varphi =
\frac{\theta_B}{\sqrt{2k}} \nonumber \\
&&T(B)^2 = \frac{w\bar w}{1+w\bar w} = \tanh^2~\frac{B}{\sqrt{2k'}}\nonumber
\\
&&C(B)^2 = 1+w\bar w = \cosh^2~\frac{B}{\sqrt{2k'}}
\label{24}
\eea
In both Eqs. (21) and (23), the pure fermionic part of the action is given
by the free-fermion kinetic terms together with some current-current
interaction. This fact permits us to describe the four left and the four
right fermions in terms of free bosonic fields $\phi_A, \phi_B$
compactified in the shifted radii \cite{ee}:
$$
\matrix{
\phantom{subspaces and us to subspaces }
R^2_A &=& 1 + \frac{2}{k}  = \frac{k+2}{k}~;~~~~
\bigg(\frac{SU(2)}{U(1)}\bigg)_k
&\phantom{valid in both and}(25a)\cr
\phantom{subspaces and us to subspaces }
R^2_B &=& 1 - \frac{2}{k'}  = \frac{k'-2}{k'}~;~~
\bigg(\frac{SL(2,R)}{U(1)}\bigg)_{k'}
&\phantom{valid in both and }(25b)}
$$
\addtocounter{equation}{1}
The deviation from the value $R_A = R_B = 1$ is due to current-current
interactions. The decoupling of $\phi_A$ and $\phi_B$ fields can be seen
via the bosonization of fermions and the redefinition of the
$\Pbar\theta_A$ and $\Pbar\theta_B$ bosonic currents order by order in a
$\frac{1}{k}$ or $\frac{1}{k'}$ expansion. This statement is indeed exact
and it follows from the fact that both $\bigg(\frac{SU(2)}{U(1)}\bigg)_k$
and  $\bigg(\frac{SL(2,R)}{U(1)}\bigg)_{k'}$ are exact $N = (2,2)$
superconformal models. This fact is well known in both $\frac{SU(2)}{U(1)}$
and $\frac{SL(2,R)}{U(1)}$ supersymmetric coset models in the compact and
non-compact parafermionic representations. The $N = 2$ generators, $J(\xi
), G(\xi ), \bar G(\xi )$ and $T(\xi )$ are given in terms of free-scalar
fields $(\phi_A,\phi_B)$ and in terms of (non-local) parafermionic currents
\cite{ss}-\cite{vv}, $(P_k,\Pi_{k'})$,
\begin{quote}
\item{~(i) - $\bigg(\frac{SU(2)}{U(1)}\bigg)_k$}
\bea
J(\xi ) &=& \sqrt{\frac{k}{k+2}}
\partial\phi_A \nonumber \\
\Gbar (\xi ) &=& \pbar_k~e^{\pm i\sqrt{\frac{k+2}{k}}\phi_A} \nonumber \\
T(\xi )& =& -\frac{1}{2}(\partial\phi_A)^2 + T_P(\xi )\phantom{xxxx}
\label{26}
\eea
\item{(ii) - $\bigg(\frac{SL(2,R)}{U(1)}\bigg)_{k'}$}
\bea
J(\xi ) &=& \sqrt{\frac{k'}{k'-2}}
\partial\phi_B \nonumber \\
\Gbar (\xi ) &=& \Pibar_{k'}~e^{\pm i\sqrt{\frac{k'-2}{k'}}\phi_B} \nonumber
\\ T(\xi ) &=& -\frac{1}{2}(\partial\phi_B)^2 + T_{\Pi}(\xi )
\label{27}
\eea
\end{quote}

$\phi_A$ and $\phi_B$ are free bosons compactified on radii $R_A =
\sqrt{\frac{k+2}{k}}$ and $R_B = \sqrt{\frac{k'-2}{k'}}$ and parametrize the
fermions $\psi^a_{\pm}, a = 1,2,3,4$ which appear in the $\sigma$-model
actions in Eqs. (21) and (23) according to our previous discussion.
For large $k, k'$, the operators $e^{i\sqrt{\frac{k+2}{k}}\phi_A}$ and
$e^{i\sqrt{\frac{k'-2}{k'}}\phi_B}$  have conformal
dimensions of almost $\frac{1}{2}~~(h_A = \frac{1}{2} + \frac{1}{k},~h_B =
\frac{1}{2} - \frac{1}{k'}$) $P_k$ and $\Pi_{k'}$ are the parafermionic
currents with dimensions $h_P = 1 - \frac{1}{k}$ and $h_{\Pi} = 1 +
\frac{1}{k'}$, so that $G(\xi )$ in Eqs. (26) and (27) are the $N = 2$
supercurrents of dimension $\frac{3}{2}$.

As usual, the compact and non-compact parafermions satisfy the algebra
which follows from their O.P.E. \cite{ss}-\cite{uu}:
\bea
P_k(\xi )\bar P_k(\xi ') &=&
2\bigg[\bigg(\frac{k}{k+2}\bigg)~\frac{1}{(\xi -\xi ')^2} + T_P(\xi
')\bigg]~(\xi -\xi ')^{\frac{2}{k}}\nonumber \\
\Pi_{k'}(\xi )\bar \Pi(\xi ') &=&
2\bigg[\bigg(\frac{k'}{k'-2}\bigg)~\frac{1}{(\xi -\xi ')^2} + T_{\Pi}(\xi
')\bigg]~(\xi -\xi ')^{-\frac{2}{k'}}
\label{28}
\eea
and the dimension-2 operators $T_P(\xi )$ and $T_{\Pi}(\xi ')$ are their
corresponding stress tensors. $T_P(\xi )$ and $T_{\Pi}(\xi ')$ satisfy the
Virasoro algebra
\beq
T(\xi )T(\xi ') = \frac{c}{2(\xi -\xi ')^4} + \frac{2T(\xi ')}{(\xi -\xi
')^2} + \frac{\partial T(\xi ')}{(\xi -\xi ')}
\label{29}
\eeq
with central charge $c_P = \frac{3k}{k+2}-1, c_{\Pi} = \frac{3k'}{k'-2}-1$.
Using the above O.P.E. [Eqs. (28), (29)] and those of the free
fields
\beq
\partial\phi_A~\partial\phi_A = -\frac{1}{(\xi -\xi ')^2}~,~~
\partial\phi_B~\partial\phi_B = -\frac{1}{(\xi -\xi ')^2}~,
\label{30}
\eeq
it is easy to show the closure of the $N = 2$ algebra for both
$\bigg(\frac{SU(2)}{U(1)}\bigg)_k$ and
$\bigg(\frac{SL(2,R)}{U(1)}\bigg)_{k'}$ supersymmetric coset models.

What is interesting is the interpretation of the almost
dimension-one, non-local cite{bkk} currents $P_k$ and $ Pi_{k'}$ for $k$
and $k'$ large cite{ee}.
$P_k$ are the conjugate momenta of the two-dimensional subspace $z$ while
$ Pi_{k'}$ are those of the $w$ subspace. As we mention in the
introduction, this interpretation breaks down for small values of $k$ and
$k'$, e.g., when the target backgrounds are strongly curved.

For large values of $k$ and $k'$, $P_k$ generalizes  two out of the four
flat space currents while $\Pi_{k'}$ generalizes the two remaining ones
\cite{ee}, \cite{bk}
\bea
\pbar_k &\buildrel{k\rightarrow\infty}\over\longrightarrow& \pm\partial x_1 +
i\partial x_2 \nonumber \\
\Pibar_{k'} &\buildrel{k'\rightarrow\infty}\over\longrightarrow& \pm\partial
x_4 + i\partial x_3
\label{31}
\eea

We would like to show that for $k' = k+4$ the $N = 2$ superconformal
symmetry is extended to $N = 4$ with $\hat c = 4 (c = 6)$ for any value of
$k$. For this purpose it is more convenient to redefine the fields $\phi_A$
and $\phi_B$ in terms of two other bosonic fields $H_+$ and $H_-$ as
follows:
\bea
\sqrt{2}H_+ &=& ~~\sqrt{\frac{k}{k+2}}~\phi_A + \sqrt{\frac{k+4}{k+2}}~\phi_B
\nonumber \\
&& \nonumber \\
\sqrt{2}H_- &=& +\sqrt{\frac{k+4}{k+2}}~\phi_A - \sqrt{\frac{k}{k+2}}~\phi_B
\label{32}
\eea
The $N = 2$ current of the $\Delta_k^{(4)}$ is then given uniquely in terms
of the $\partial H_+$ current. It is then possible (due to the $N = 2$
algebra) to factor the $H_+$ dependence from the (complex) supercurrent. One
finds \cite{ee}
\beq
G = \bigg[ P_k~e^{i\frac{\alpha}{\sqrt{2}} H_-} +
\Pi_{k'}~e^{-i\frac{1}{\alpha\sqrt{2}}H_-}\bigg]~e^{i\frac{1}{\sqrt{2}}H_+}
\label{33}
\eeq
with $\alpha = \sqrt{\frac{k+4}{k}}$.

It is suggestive to compare the above expression of the supercurrent with
that of the free-field realization \cite{ff} $(k\rightarrow\infty, \alpha =
1)$
\beq
G_{free} = \bigg[(\partial x_1 + i\partial x_2)e^{i\frac{1}{\sqrt{2}}H_-} +
(\partial x_4 + i\partial x_3)e^{-i\frac{1}{\sqrt{2}}H_-}\bigg]~
e^{i\frac{1}{\sqrt{2}}H_+}
\label{34}
\eeq
Here $H_+$ and $H_-$ are given in terms of free fermions via the
conventional bosonization
\bea
\sqrt{2}H_+ &=& \psi_1\psi_2 + \psi_4\psi_3 \nonumber \\
\sqrt{2}H_- &=& \psi_1\psi_2 - \psi_4\psi_3
\label{35}
\eea

There are some similarities and some basic differences between the two
realizations. First, the r\^ole of the $H_+$ field is the same and, as  is
expected from the $N = 2$ algebra, is a free field compactified on a
radius $R_{H^+} = \sqrt{\frac{c}{3}} = \sqrt{2}$. This value of the radius
is special and extends in both cases the $U(1)$ to $SU(2)_1$ current
algebra. In free field realization, however, the $N = 2$ superconformal
symmetry is extended to an $N = 4$, due to the existence of an additional
supercurrent, namely:
$$
\tilde G_{free} = \bigg[(\partial x_1-i\partial
x_2)e^{-i\frac{1}{\sqrt{2}}H_-} -
(\partial x_4-i\partial
x_3)e^{i\frac{1}{\sqrt{2}}H_-}\bigg]~e^{i\frac{1}{\sqrt{2}}H_+}
$$

In a similar way,
an additional supercurrent exists also in the $\Delta_k^{(4)}$ realization
\cite{ee}: \beq
\tilde G = \bigg[ \bar P_k~e^{-i\frac{\alpha}{\sqrt{2}} H^-} -
\bar\Pi_{k'}~e^{i\frac{1}{\alpha\sqrt{2}}H^-}\bigg]~e^{i\frac{1}{\sqrt{2}}H^+}
\label{36}
\eeq
The $N = 4$ superconformal algebra closes in both realizations among $[T_B,
G, \tilde G$ and $S_i]$ with central charge $\hat c = 4$. This shows that the
$\Delta_k^{(4)}$ space shares the same global superconformal symmetries as
the $T^4/Z_2$ orbifold model, the four-dimensional Calabi-Yau
space $K_3$ and the previously discussed $W_k^{(4)}$ wormhole space.
$\Delta_k^{(4)}$ and $W_k^{(4)}$  are not only  non-trivial spaces but are
also non-compact spaces ($K_3$ and $T^4/Z_2$ are compact).

For completeness, we will present below the basic operators and fields of
the $W_k^{(4)}$ realization
\cite{ggg},\cite{hh},\cite{bb}-\cite{ee},\cite{kk}. As we already
mentioned, $W_k^{(4)}$ is based on a supersymmetric $SU(2)_k\times U(1)_Q$
WZW model with a background term $Q = \sqrt{\frac{2}{k+2}}$ in the $U(1)_Q$
current. The four fermions of the model are free and are parametrized by
the $H^+$ and $H^-$ fields (via bosonization) as in the free-field
representation [Eq. (35)]. The four coordinate currents are the three
$SU(2)_k~~ (J^i, i = 1,2,3)$ currents and the $U(1)_Q ~~J_4$ current
\bea
J^i(\xi )J^i(\xi ') &=& -\frac{k}{2}~\frac{\delta^{ij}}{(\xi -\xi ')^2} +
\epsilon^{ij\ell}~\frac{J^{\ell}}{(\xi -\xi ')^2}~,~~i = 1,2,3\nonumber \\
J^4(\xi )J^4(\xi ') &=& \frac{-1}{(\xi -\xi ')^2} \cdot \frac{1}{Q^2}
\label{37}
\eea
(the $Q$ rescaling of $J^4$ is for convenience).

The $T_B, B, \tilde G$ and $S_i$ associated to $W_k^{(4)}$ are
\bea
T_B &=& -\frac{1}{2}~\bigg[ (\partial H^+)^2 + (\partial H^-)^2 + Q^2~~
(J^2_1+J^2_2+J^2_3+J^2_4+\partial J_4)\bigg]
\nonumber \\
G &=& Q~\bigg[(J_1+iJ_2)~e^{i\frac{1}{\sqrt{2}}H^{-}} +
(J_4 +i(J_3+\sqrt{2}\partial H^-))~\bar e^{i\frac{1}{\sqrt{2}}H^-}\bigg]
{}~e^{i\frac{1}{\sqrt{2}}H^+}\nonumber \\
\tilde G &=& Q~\bigg[(J_1-iJ_2)~\bar e^{i\frac{1}{\sqrt{2}}H^{-}} -
(J_4 - i(J_3+\sqrt{2}\partial H^-))~e^{i\frac{1}{\sqrt{2}}H^-}\bigg]
{}~e^{i\frac{1}{\sqrt{2}}H^+}\nonumber \\
S_0 &=& \frac{1}{\sqrt{2}} \partial H^+~,~~S_{\pm} = e^{\pm i\sqrt{2}H^+}
\label{38}
\eea
In the $W_k^{(4)}$ realization, both $H^+$ and $H^-$ are compactified on
a torus with radius $R_{H^+} = R_{H^-} = \sqrt{2}$ as in the free field case.
There are in total three underlying $SU(2)$ Kac-Moody currents: (i) the
$SU(2)_k$ defined by the coordinate currents, (ii) the $SU(2)^+_1$ defined
by the $H^+$ field $R_{H^+} = \sqrt{2}$, (iii) the $SU(2)^-_1$ defined by the
$H^-$ field $R_{H^-} = \sqrt{2}$.

The background term in $T_B, (Q\partial J^4)$, comes from a non-trivial
dilaton background: $\Phi = Q X^4$. The term
$Q\sqrt{2}\partial H^-~exp[\pm i\frac{1}{\sqrt{2}}H^- + i\frac{1}{\sqrt{2}}
H^+]$
in $G$ and $\tilde G$
describes, at the same time, the standard fermionic torsion term $\pm
Q\psi^i\psi^j\psi^k ~~(i = 1,2,3,4)$ as well as the fermionic background term
$\pm Q\partial\psi^i$. They are arranged together in $\sqrt{2}\partial H^-$
which shifts the $J_3$ current. Contrary to the $\Delta_k^{(4)}$ space in
which the torsion terms were absent, in $W_k^{(4)}$ space there is a
non-vanishing torsion due to the WZ term proportional to the $SU(2)$
structure constant. So, in $W_k^{(4)}$ the antisymmetric field
background $B_{ij}$ is non-zero with non-trivial field strength $H_{ijk}$
\beq
H_{ijk} = e^{-2\phi}\epsilon_{ijk}~{^{\ell}\partial_{\ell}}\Phi \simeq
e^{-2\phi}Q\cdot \epsilon_{ijk}^{\phantom{ijk}4}
\label{39}
\eeq

Summarizing our results, we propose a special class of exact string
solutions \cite{bb}-\cite{ee} with $N = (1,0)$ or $N = (1,1)$ local
worldsheet supersymmetry. In the type II case, we arrange the degrees of
freedom of the ten supercoordinates in three superconformal systems:
$$
\hat c = 10 = \{\hat c = 2 \}_0 + \{\hat c = 4\}_1 + \{\hat c = 4\}_2
$$
The $\hat c=2$ system is saturated by two free superfields (compact or
non-compact), and so the background metric is flat:
$$
ds^2\{F^2\} = dxd\bar x~,~~(x = x_1+ix_2)~.
$$
The remaining eight supercoordinates appear in a group of four. In both
$\{\hat c = 4\}_{1,2}$ subsystems an $N = 4$ globally defined superconformal
symmetry is assumed. We propose three exact $N = 4$ superconformal systems
with $\hat c = 4$.

\begin{quote}
\item{~~(i) - $Flat~realization : \hat c = 4$}
$$
\matrix{
&ds^2(F) = dzd\bar z + dwd\bar w \hfill\cr
& \cr
&\Phi = {constant}~, ~~ B_{ij} = {constant}\cr}
$$

This realization includes the flat space as well as the toroidal $T^4$ and
orbifold $T^4/Z_2$ models.
\vspace*{0.5cm}
\item{~(ii) - $Semi-wormhole~space~W_k^{(4)}~with~ \hat c =
4$}
$$
ds^2(W) = k~\frac{dzd\bar z + dwd\bar w}{z\bar z + w\bar w}$$
$$2\Phi = \log (z\bar z + w\bar w)$$
$$H_{ijk} = \epsilon_{ijk\ell}\partial^{\ell}\Phi
$$
\item{(iii) - $The ~\Delta_k^{(4)} realization~with~ \hat c =
4$}
$$
ds^2(\Delta ) = k~\frac{dzd\bar z}{1-z\bar z} + (k+4)~\frac{dwd\bar
w}{w\bar w + \epsilon}$$
$$ 2\Phi(\Delta ) = \log (1-z\bar z) + \log (w\bar w + \epsilon)~,~~\epsilon
= \pm 1$$
$$H_{ijk} = 0$$
$$\epsilon = 1~,~~{\rm Bell-Cigar~space}$$
$$\epsilon = -1~,~~{\rm Bell-Trumpet~space}$$
\end{quote}
\vspace*{0.5cm}

	Using as building blocks the $F^2, F^4, \Delta^4_{\epsilon}$ or $W^4$
superconformal systems, one obtains several string solutions based on
non-trivial ten-dimensional backgrounds. The advantage of our solutions
compared to the subclass of models $(F^6\times W^{(4)})$ studied in the
literature \cite{hh} lies in the complete knowledge of the $\{\hat c =
2\}_0$ and $\{\hat c = 4\}_A$ superconformal theories. Due to this we are
able to study not only the background solutions for large $k_A$ but also
the full string spectrum and derive the partition functions of the models
in a modular invariant way respecting the $N = 4$ superconformal symmetry
\cite{bb}-\cite{ee}. Indeed, the full spectrum is given in terms of several
characters of known conformal theories. In the $W^4_k$ realization
\cite{bb}-\cite{dd} one uses a character combination of $SU(2)_k, SU(2)^+_1$
and $SU(2)^-_1$ together with the $U(1)_Q$ Liouville-type characters. In
$\Delta^{(4)}_k$, one \cite{ee} uses the compact \cite{kp} and non-compact
\cite{bbk}
parafermionic characters (string functions) together with the $SU(2)^+_1$ and
$U(1)^-_{R_{H^-}}~~~ (R_H = \sqrt{2}~\sqrt{\frac{k+4}{k}})$. Finally in $F^2$
and $F^4$, one uses non-compact, toroidal or orbifold characters.

	It is important to stress here that the character combinations above, are not
arbitrary but  are dictated by the global existence of $N = 4$
superconformal symmetry as well as modular invariance.

These requirements define some generalized GSO projections
\cite{bb}-\cite{ee}, similar to that of the fermionic construction
\cite{ww} and that of the conformal block construction \cite{yy} of Gepner.
One of these projections is fundamental and guarantees the existence of
some space-time supersymmetries via the $N = 4$ spectral flows. It can be
expressed in terms of the two $N = 4 ~~~(\hat c = 4)_A, A =
1,2~~\{SU(2)^+_1\}_A$ spins, $[j(s_1), j(s_2)]$ \cite{bb}-\cite{ee}:
\beq
2(j(s_1) + j(s_2)) = {\rm odd~integer}~.
\label{40}
\eeq

Equation (\ref{40}) guarantees the existence of some covariantized target
space symmetries around the proposed non-trivial backgrounds. This
stabilizes our solutions (at least perturbatively) and projects out all
kinds of tachyonic or complex conformal weight states. This projection
phenomenon is similar to the one observed by Kutasov and Seiberg in the
framework of non-critical superstrings with an $N = 2$ globally defined
superconformal symmetry \cite{zz}.

Heterotic solutions are simply obtained via a generalized
\cite{aai},\cite{yy} heterotic map. Here also the solutions are stable but
the number of covariantized space-time supersymmetries is reduced by a
factor of two.

	We hope that our explicit construction of a family of consistent and
stable solutions will give a better understanding of some fundamental
string properties, especially in the case of strongly curved backgrounds
(small $k_A$) where the notion of the space-time dimensionality and
topology breaks down.

\vfill
\eject
\vspace*{1cm}
\noindent
{\bf ACKNOWLEDGEMENTS}

I would like to thank I. Antoniadis and S. Ferrara. Some results about the
$W_k^{(4)}$ space of this work are products of our collaboration. I am also
grateful to C. Callan, D.~Gross, E. Kiritsis and D. L\"ust for fruitful
discussions.

 \end{document}